\documentclass[12pt,preprint]{aastex6}
\usepackage{graphicx}
\usepackage{textcomp}
\usepackage{natbib}
\usepackage{subfigure}
\usepackage{amsmath}
\usepackage{amssymb}
\usepackage{hyperref}
\usepackage{multirow}\usepackage[section] {placeins}

\begin{document}

\title{An Equation of State of CO for use in Planetary Modeling}

\author{M. Podolak}
\affiliation{Dept. of Geosciences, Tel Aviv University, Tel Aviv, 69978 Israel}

\author{A. Levi}
\affiliation{Braude College of Engineering, Karmiel, 2161002 Israel}
\email{amitlevi.planetphys@gmail.com}

\author{A. Vazan}
\affiliation{Astrophysics Research Center (ARCO), Dept. of Natural Sciences, Open University of Israel, Raanana, 43107 Israel}

\author{U. Malamud}
\affiliation{Dept. of Geosciences, Tel Aviv University, Tel Aviv, 69978 Israel \\
 Department of Physics, Technion – Israel Institute of Technology, Technion City, 3200003 Haifa, Israel}


\section*{ABSTRACT}
Although carbon monoxide (CO) is an abundant molecule and may have great importance for planetary interiors, measurements of its properties are difficult due to its extreme volatility.  We calculate the equation of state for CO over a range of temperature and density that is applicable to the conditions in planetary interiors. Previous experimental and theoretical studies cover only a limited temperature-density range.  Our calculations match these early results well, but now cover the full range of relevance. The method of calculation is based on the general-purpose quotidian equation of state described by \cite{more88}, which is here used in order to generate a freely downloadable look-up table to be used by the community.

\section{INTRODUCTION}
When modeling planetary interiors, it is necessary to have adequate descriptions for the behavior of the constituent materials. Thus equation of state (EOS) tables have been produced for the two most abundant elements in the universe, hydrogen and helium \citep[see, e.g.][]{chabrier2019}, as well as other materials expected to be of importance for planet models, such as water \citep[see, e.g.][]{haldemann2020}, various silicates such as dunite \citep{benz1989}, granite \citep{PierazzoEtAl-1997}, basalt \citep{PierazzoEtAl-2005}, quartz \citep{Melosh-2007} and important metals such as iron \citep[e.g.][]{emsenhuber2018}.

Since both carbon and oxygen have relatively high cosmic abundances, and since CO is a very stable molecule, CO could be an important constituent in planetary interiors \citep[see, e.g.][]{LisseEtAl-2022}. Yet this possibility cannot be properly addressed because only limited regions of the CO EOS have been studied, and there are no complete equation of state tables available in the literature. Empirical measurements of the density of solid ($\alpha$-cubic, $\beta$-hexagonal) and liquid CO have been made \citep{BoonEtAl-1967,Bierhals-2001}, in addition to various other physical properties such as viscosity, heat capacity \citep{RudeskoSchubnikow-1934,TancrediEtAl-1994}, and elastic constants \citep{gammon1978}. All of these studies are applicable to extremely low temperature and pressure conditions, and are ill-suited for planetary interior applications. The behavior of CO at higher pressures and temperatures has been studied, to a limited extent by \cite{nellis1981} who reported the results of shock experiments.  More recent work by \cite{zhang2011} gives a more refined hugoniot for CO.  In addition, theoretical calculations by \cite{goodwin1985} have investigated the region of pressures below 100\,MPa.  Individual pressure-temperature-density points have been computed from quantum molecular dynamics calculations by \cite{massacrier2011}, \cite{wang2010}, and \cite{leonhardi2017}.  However, all of this data is insufficient for planetary modeling, where a much larger range of pressures and temperatures are encountered.

The fact that shock-derived carbon condensates have diameters of the order of a few nanometers \citep{titov1989, viecelli2001, kruger2005}, and growth timescales of 100's of picoseconds \citep{armstrong2020}) make direct DFT based molecular dynamics simulations of this system particularly challenging. Overcoming such immense difficulties often requires some synthesis between a DFT based approach and more classical force field models using various training models often referred to as machine learning approaches \citep[see, e.g.][]{lindsey2020, singraber2019}. These techniques are very demanding computationally. Therefore, our model which is in good agreement with experimental data and covers a very wide pressure-temperature domain is of merit.
	
To this end we have generated an equation of state table for CO which we describe below.  Our calculation is admittedly more crude, but it should be sufficiently close to reality so as to be useful in establishing model trends such as was done in the models of \cite{podolak2022}, for example.  This paper is structured as follows: Section 2 gives a brief description of the method for computing the quotidian EOS (QEOS).  This computation requires the knowledge of the density and bulk modulus at low energy.  The DFT calculation of these parameters is described in section 3, and the results are given in section 4.  The resulting EOS table and its comparison to experimental and theoretical work described above is given in section 5.  It is hoped that this work will encourage more detailed EOS modeling for CO in the future.

\section{Quotidian Equation of State}

\cite{more88} present a general-purpose method for computing equations of state at high pressure, called the Quotidian Equation of State (QEOS). The QEOS is a statistical-mechanics-based method, in which thermodynamic quantities are derived from the Helmholtz free energy. The Helmholtz free energy term is composed of three parts: an ionic contribution, an electronic contribution, and a bonding correction. 
The ionic part is calculated by the Cowan model, a semi-empirical model which interpolates between known limiting physical cases (ideal gas law, Lindemann melting law, Dulong-Petit law, Gr\"uniesen EOS, Debye lattice).
The electronic part is calculated using a modified Thomas-Fermi (TF) model. The TF model neglects attractive (bonding) forces between neutral atoms and therefore overestimates the critical point and the pressure near normal conditions. 
The bonding correction is used here to correct for the electronic part failure by calibration of the EOS with density and bulk modulus at reference conditions of zero (low) energy. 

This method has been used to develop EOS tables for Fe, SiO$_2$ and H$_2$O for use in planetary modeling which compare well with other EOS tables such as SESAME and ANEOS for these substances \citep{vazan13, vazan2018, vazan2022}. 
The QEOS input variables are: atomic number, atomic weight, and reference conditions density and bulk modulus. The calculated quantities are: pressure, specific internal energy, and specific entropy.  The temperature-density range of the calculation is  $11.6<T<1.16\times 10^6$\,K, and $2.5\times 10^{-13}<\rho<100$\, g\,cm$^{-3}$.
The liquid-vapor phase transition is determined with regard to the Maxwell construction, based on finding equal Gibbs free energy on the liquid and the vapor sides of each isotherm (up to the critical temperature). As a result, there is no coexistence of vapor and liquid phases in the resulting smooth QEOS.  

In order to calculate a QEOS for CO, the method requires prior knowledge of the density and bulk modulus of the material at very low temperature and pressure. Unfortunately there have been no measurements of these quantities for the $\alpha$-phase of CO.  We therefore performed a first-principles calculation for this state using density-functional theory (DFT). This calculation described in the next section.

\section{COMPUTATIONAL METHODS}\label{COMPUTATIONAL METHODS}

Here we study the equation of state of $\alpha$-CO at $0$\,K. The structure is taken from \cite{Hall1976}.  We performed static total energy relaxations with the CP2K code \citep{Thomas2020}. We use the quickstep framework within CP2K with the Gaussian and plane waves mixed bases (GPW). We adopt the Gaussian basis sets from \cite{VandVondele2005, VandeVondele2007},  in conjunction with the pseudopotentials (GTH-PBE) of Goedecker, Teter, and Hutter \citep{Goedecker1996,Hartwigsen1998,Krack2005}.

Our system is converged for a planewave cutoff energy of $600$\,Ry and a REL\_CUTOFF of $40$\,Ry.  
We use the revised PBE exchange functional GGA\_X\_PBE\_R from \cite{Zhang1998} and a PBE correlation functional, GGA\_C\_PBE \citep{Perdew1996,Perdew1997}. These are found to be adequate choices when describing an aqueous system in conjunction with the non-local van der Waals correlation using the Grimme D3 method \citep{Grimme2010}, achieving convergence for R\_CUTOFF of $14$. The calculations were done done on a 2x2x2 supercell consisting of $32$ CO molecules.  The derived data at $0$\,K is obtained using CELL\_OPT within CP2K and reported below.  

\section{The Equation of State}

\begin{figure}[ht]
\centering
\includegraphics[trim=0.15cm 5cm 0.01cm 5.0cm , scale=0.55, clip]{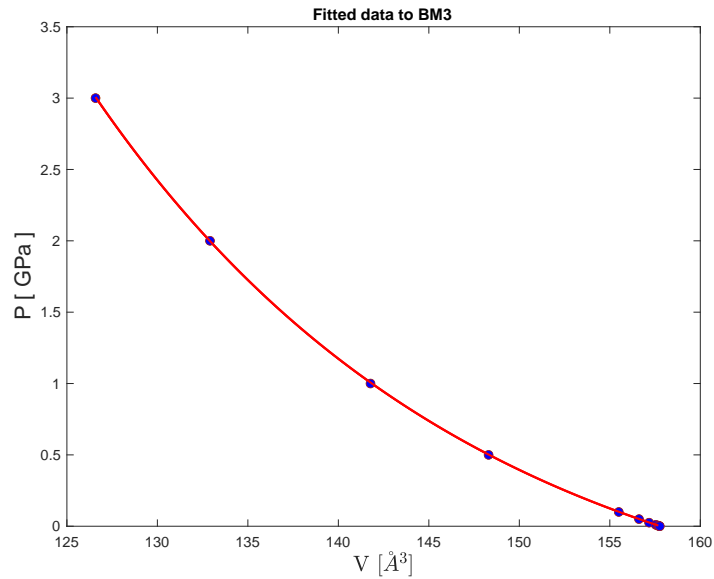}
\caption{\footnotesize{Pressure versus unit cell volume for $\alpha$-CO. The blue circles are unit cell volumes from our optimization data at $0$\,K, and the solid red curve is the fitted third order Birch-Murnaghan equation of state (BM3).  }}
\label{fig:BM3EOS}
\end{figure}

In table \ref{tab:EOSdata} and fig.\,\ref{fig:BM3EOS} we give the volumes and energies derived for different pressures at $0$\,K. This data is fitted to a third order Birch-Murnaghan equation of state with a bulk modulus B$=6.556\pm0.074$\,GPa, a pressure derivative for the bulk modulus of B$'=6.846\pm0.120$, and a zero pressure volume of V$_0=157.80\pm0.05$\,\AA$^3$. The error bars are at the $2\sigma$ level.  As mentioned above, the QEOS requires a knowledge of $\rho$ and B at reference conditions of zero energy, and, based on this calculation, and a fit to the four lowest pressure points we take $\rho=1.179$\,g cm$^{-3}$ and B$=2.676$\,GPa as the input parameters.  Note that this value of B falls between the best fit value of 6.556\,GPa given above and the value of 1.3\,GPa measured by \cite{gammon1978} for $\beta$-CO.

\begin{deluxetable}{cccc}
\tablecolumns{4}
\tablewidth{0pt}
\tablenum 1
\tablecaption{The volume, internal energy, and derived enthalpy as a function of pressure for the $\alpha$-CO solid. Data is for a cubic supercell consisting of $32$ CO molecules.}
\tablehead{
  \colhead {$P$}    & \colhead{$V$}  & \colhead{$U$}  & \colhead{$H$}   \\
  \colhead{$[bar]$}  & \colhead{$[\AA^3]$} & \colhead{$[Ha]$}  & \colhead{$[Ha]$}  }
\startdata
30,000         &     1012.664     &     -694.3516      &    -693.6548     \\ 
\hline
20,000         &     1063.250     &     -694.3819     &     -693.8941    \\ 
\hline
10,000         &      1134.213    &      -694.4058    &    -694.1456    \\ 
\hline
5000            &     1186.408     &     -694.4146      &    -694.2785    \\ 
\hline
1000            &     1243.957    &     -694.4184    &    -694.3899    \\ 
\hline
500             &     1252.903     &     -694.4185     &     -694.4041    \\ 
\hline
250             &     1257.361     &      -694.4186      &    -694.4114     \\ 
\hline
100             &     1260.298     &      -694.4186      &    -694.4157    \\ 
\hline
50               &     1261.144     &     -694.4186     &    -694.4172     \\ 
\hline
25               &     1261.663    &     -694.4186      &    -694.4179    \\ 
\hline
10               &     1261.931     &      -694.4186      &    -694.4183     \\ 
\hline
1                 &     1262.103     &     -694.4186      &    -694.4186     \\   
\enddata
\label{tab:EOSdata}
\end{deluxetable}

Using the results of the DFT calculation described above in the quotidian code, we produced an equation of state table giving the pressure, energy and entropy of CO for a large range of temperatures and densities.  

\section{Comparison to other results}
\begin{figure}[ht]
	\centering
	\includegraphics[width=18cm]{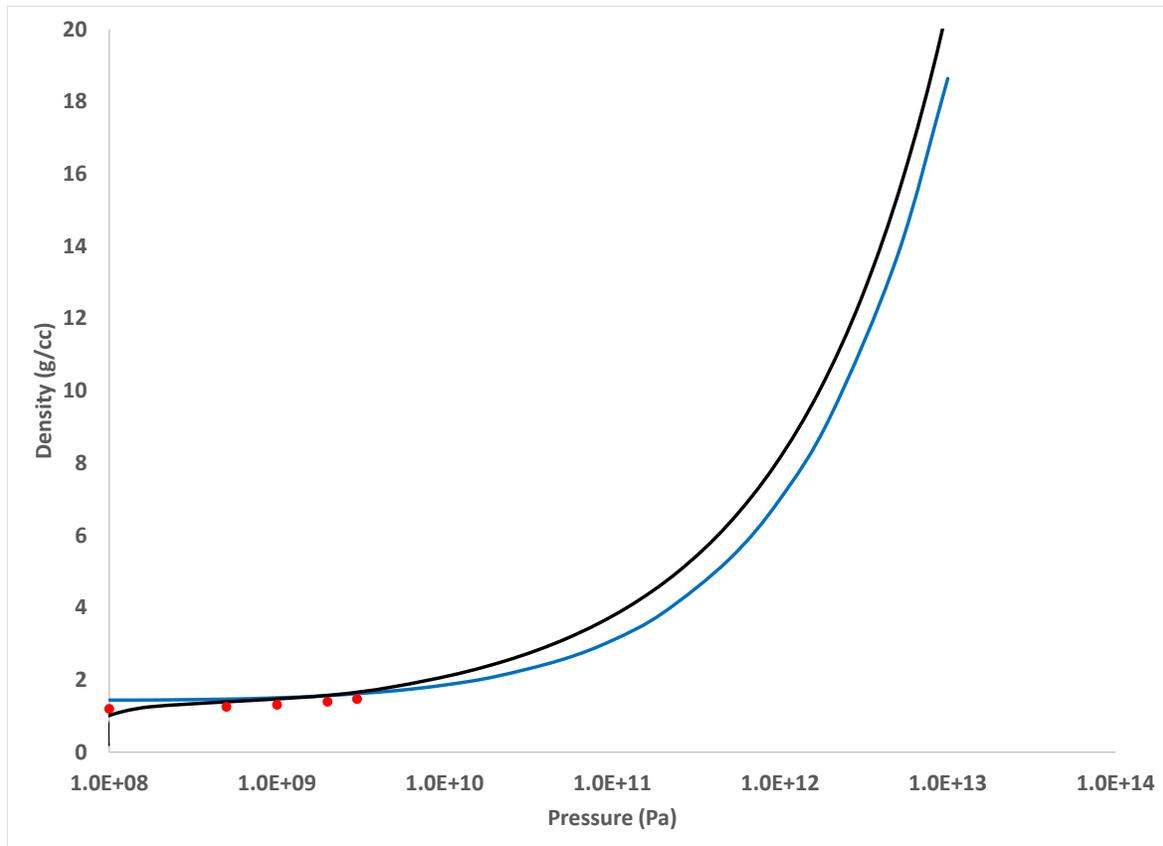}
	\caption{\footnotesize{Density as a function of pressure at zero temperature for the quotidian equation of state (black curve), and for the S-Z equation of state (blue curve).  The red dots are the results of the DFT calculation.  }}
	\label{fig:szcomp}
\end{figure} 

\cite{salpeter1967} (S-Z) describe a semi-empirical formula for predicting the zero temperature pressure-density relation for materials with any average atomic number.  In principle, the S-Z EOS is similar to the \cite{more88} approach, since it relies on a Thomas-Fermi-Dirac model of the atom.  However it does not include the effect of temperature, so it is not always suitable for planet modeling.  Fig.\,\ref{fig:szcomp} shows the comparison between our quotidian equation of state (QEOS) at zero temperature, and the S-Z EOS.  As can be seen, the agreement is excellent, and improves at higher pressures, as expected.  The red dots in the figure are the DFT calculations given in table\,\ref{tab:EOSdata} and fig.\,\ref{fig:BM3EOS}.  These fall right on the QEOS curve.  

\begin{figure}[ht]
	\centering
	\includegraphics[width=18cm]{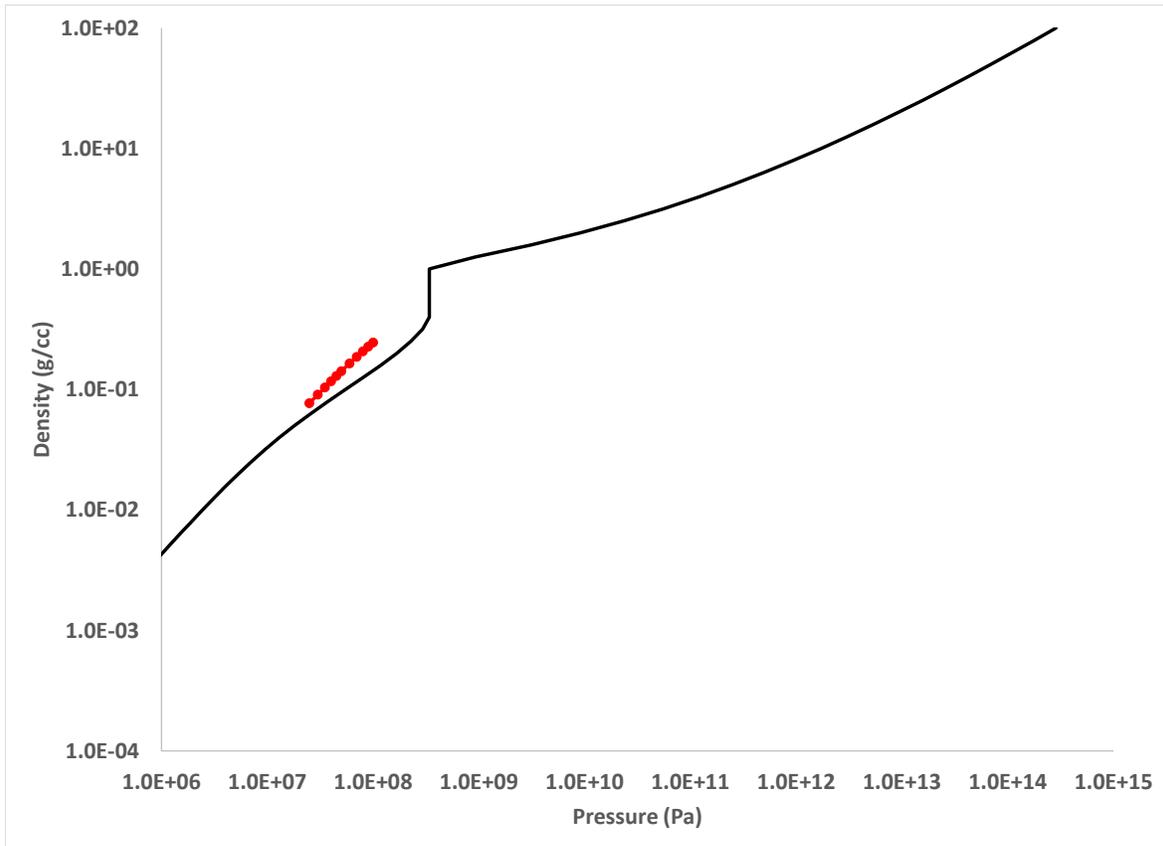}
	\caption{\footnotesize{Density as a function of pressure for an isotherm at $T=1000$\,K (black curve), compared to the data in \cite{goodwin1985} (red dots).  See text for details.  }}
	\label{fig:goodwincomp}
\end{figure}

\begin{figure}[ht]
	\centering
	\includegraphics[width=18cm]{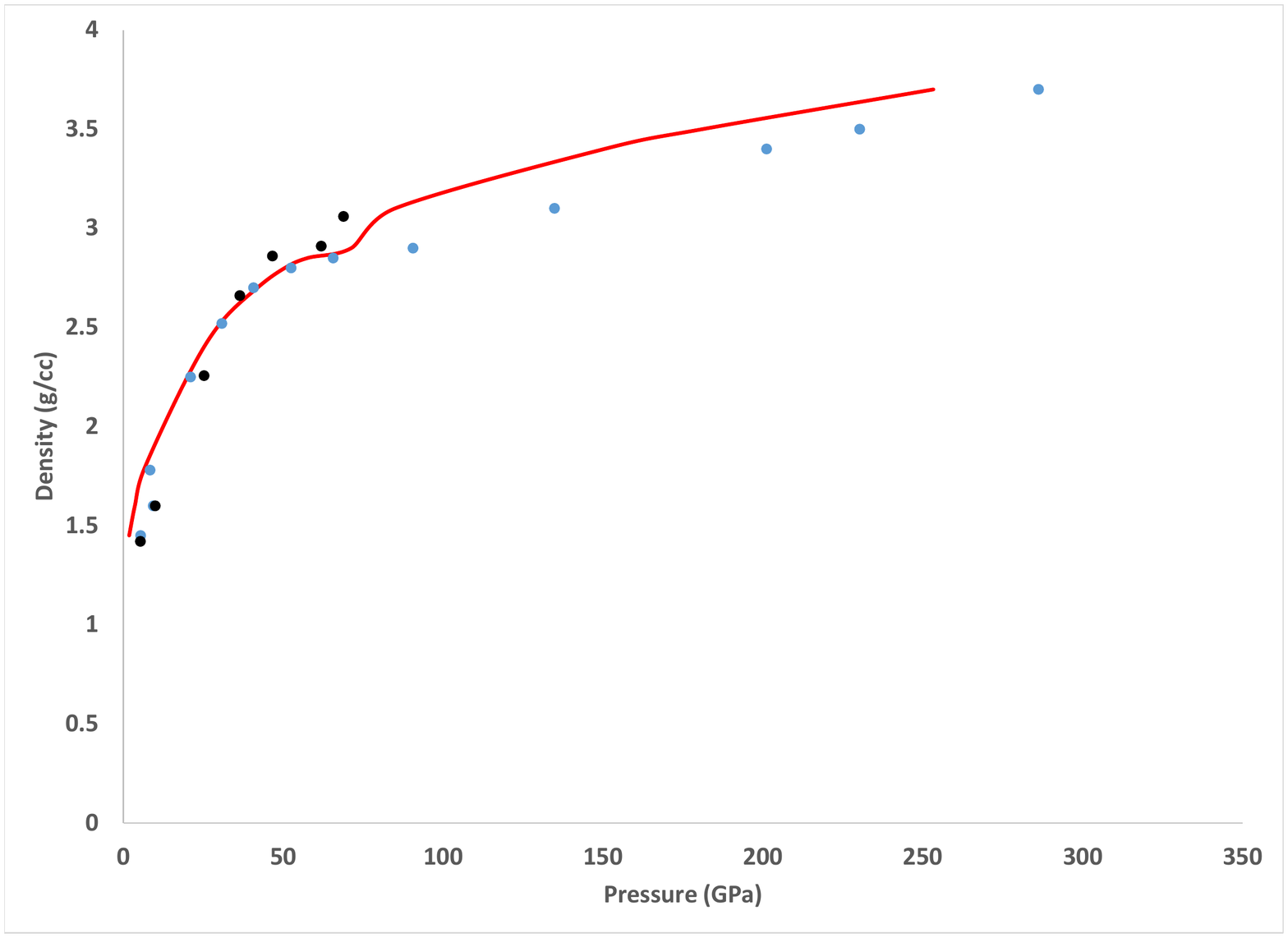}
	\caption{\footnotesize{Density as a function of pressure for a hugoniot (blue curve) corresponding to the conditions of the shock experiments of \cite{nellis1981} (black dots) and the quantum molecular dynamics calculations of \cite{zhang2011} (blue dots).   }}
	\label{fig:nelliscomp}
\end{figure} 

The QEOS can be compared to experimental data at higher temperatures as well.  \cite{goodwin1985} gives the thermophysical properties of CO up to a pressure of 100\,MPa.  Fig.\,\ref{fig:goodwincomp} shows that data for an isotherm at 1000\,K (red dots) compared to the QEOS isotherm at that temperature (black curve).  The discontinuity in the QEOS is due to the fact that the QEOS finds two phases in present in this pressure-temperature range and traverses this region using a Maxwell construction.  As a result, the computed pressure remains constant over the relevant density range.  The actual pressure, as shown by the red dots, increases along the extrapolation of the lower part of the curve, as expected. The exact position of the phase transition is sensitive to the choice of input parameters (zero energy density and bulk modulus), and the actual value may be shifted somewhat.

At still higher pressures and temperatures, there are the shock wave experiments of \cite{nellis1981}.  In this case the temperatures are only inferred from the Hugoniot relations, and are different for the different pressures.  More recently, \cite{zhang2011} have used quantum molecular dynamics calculations to compute points along a hugoniot.  These are shown (blue dots) together with the hugoniot calculated from our QEOS in Fig.\,\ref{fig:nelliscomp}.  The black dots are the experimental points of \cite{nellis1981}.  As can be seen, the agreement is quite good and is in the range of these works.  At the highest temperatures ($T\gtrsim 10^5$\,K) dissociation and ionization become important, and these effects are not directly included in our calculation.  Nonetheless, the energies we compute for CO at $T=5\times 10^5$\,K for densities of 0.1, 1, 10, and 100 g\,cm$^{-3}$ all fall within a factor of 1.5 or less from the values shown in fig. 9 of \cite{massacrier2011}. 

\begin{figure}[ht]
	\centering
	\includegraphics[width=18cm]{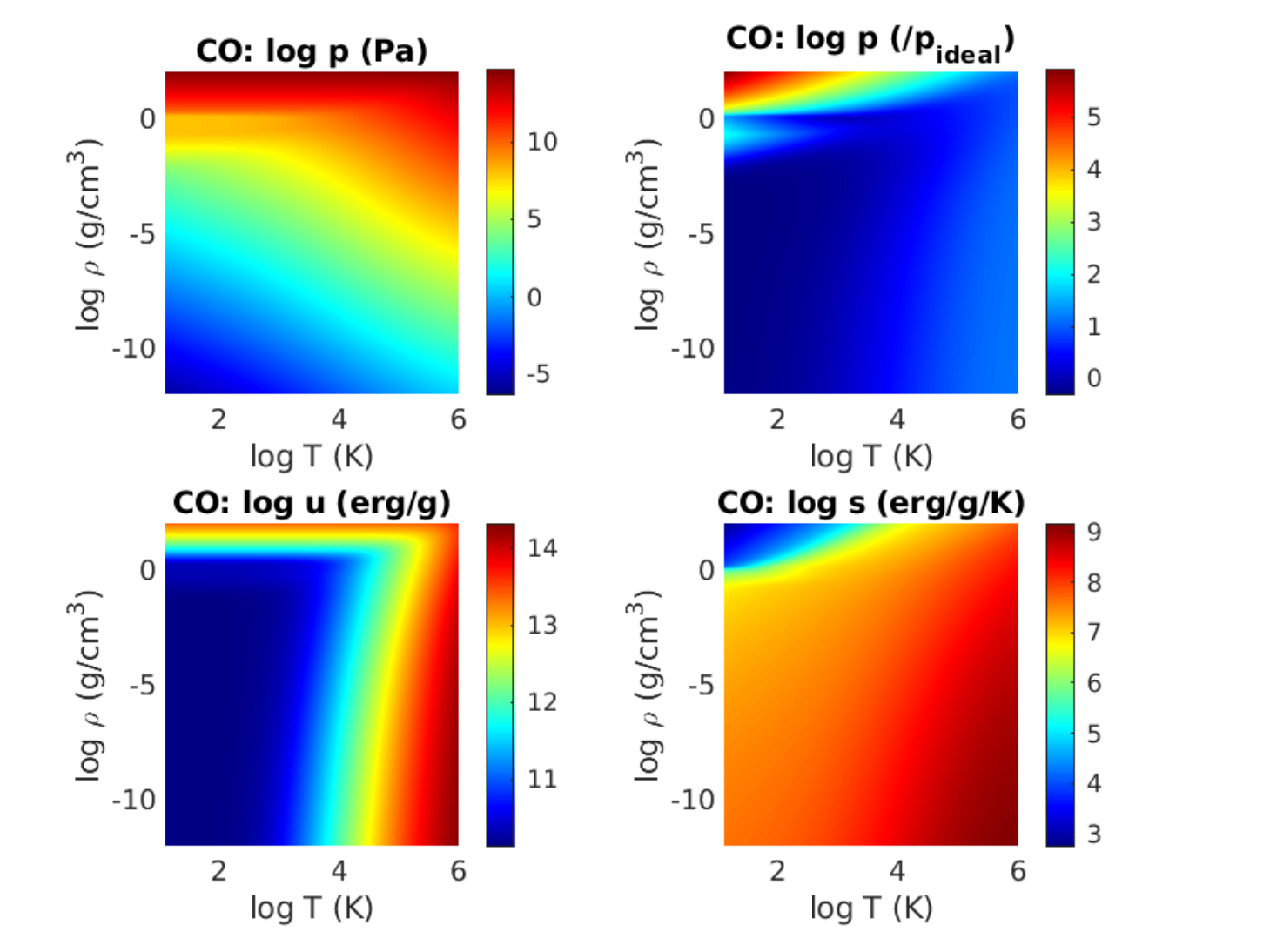}
	\caption{\footnotesize{Thermodynamic properties of CO as a function of density and temperature as computed from the quotidian equation of state. {\bf Upper left:} total pressure. {\bf Upper right:} pressure divided by ideal gas pressure.  This shows the region where an ideal gas approximation may be used.  {\bf Lower left:} specific internal energy.  {\bf Lower right:} specific entropy.  }}
	\label{fig:3d}
\end{figure} 

The full QEOS is summarized in Fig.\,\ref{fig:3d}.  A short version for a range of pressures and temperatures that are expected to be important for planetary interior modeling given in table \ref{tab:COEOS}, while the complete table is available at the following site: \href{https://github.com/UriMalamud/CO_EOS}{CO EOS download}.

\newpage

\begin{deluxetable}{ccccc}{h}
	\tablecolumns{5}
	\tablewidth{0pt}
	\tablenum 2
	\tablecaption{Equation of state for CO. }
	\tablehead{
	\colhead {$\log T$}    & \colhead{$\log \rho$}  & \colhead{$\log P$}  & \colhead{$\log u$} & \colhead{$\log s$}  \\
	\colhead{$[K]$}  & \colhead{$[g/cc]$} & \colhead{$[Pa]$}  & \colhead{$[erg/g]$} & \colhead{$[erg/g-K]$} }
	\startdata
1.06465		&	0.10	&		 8.25060	&		10.33294	&		4.12858  \\
1.06465		&	0.20	&		 9.33277	&		10.36235	&		3.94748  \\
1.06465		&	0.30	&		 9.90369	&		10.46041	&		3.83167  \\
1.06465		&	0.40	&		 10.35107	&		10.63593	&		3.74763  \\
1.06465		&	0.50	&		 10.73560	&		10.86075	&		3.67754  \\
1.06465		&	0.60	&		 11.08026	&		11.10070	&		3.61334  \\
1.06465		&	0.70	&		 11.39668	&		11.33587	&		3.55157  \\
1.06465		&	0.80	&		 11.69173	&		11.55846	&		3.49070  \\
1.06465		&	0.90	&		 11.96991	&		11.76663	&		3.43006  \\
1.06465		&	1.00	&		 12.23437	&		11.96089	&		3.36933  \\
1.06465		&	1.10	&		 12.48741	&		12.14250	&		3.30834  \\
1.06465		&	1.20	&		 12.73083	&		12.31285	&		3.24704  \\
1.06465		&	1.30	&		 12.96602	&		12.47327	&		3.18540  \\
1.56465		&	0.10	&		 8.25291	&		10.33305	&		5.29718  \\
1.56465		&	0.20	&		 9.33286	&		10.36239	&		4.92635  \\
1.56465		&	0.30	&		 9.90370	&		10.46043	&		4.63269  \\
1.56465		&	0.40	&		 10.35107	&		10.63594	&		4.42170  \\
1.56465		&	0.50	&		 10.73560	&		10.86075	&		4.27449  \\
1.56465		&	0.60	&		 11.08026	&		11.10070	&		4.16723  \\
1.56465		&	0.70	&		 11.39668	&		11.33587	&		4.08205  \\
1.56465		&	0.80	&		 11.69173	&		11.55846	&		4.00839  \\
1.56465		&	0.90	&		 11.96991	&		11.76663	&		3.94060  \\
1.56465		&	1.00	&		 12.23437	&		11.96089	&		3.87576  \\
1.56465		&	1.10	&		 12.48741	&		12.14250	&		3.81236  \\
1.56465		&	1.20	&		 12.73083	&		12.31285	&		3.74960  \\
1.56465		&	1.30	&		 12.96602	&		12.47327	&		3.68706  \\
2.06465		&	0.10	&		 8.32026	&		10.33617	&		6.33748  \\
2.06465		&	0.20	&		 9.33736	&		10.36427	&		6.10876  \\
2.06465		&	0.30	&		 9.90446	&		10.46124	&		5.82885  \\
2.06465		&	0.40	&		 10.35123	&		10.63621	&		5.53226  \\
2.06465		&	0.50	&		 10.73564	&		10.86083	&		5.24767  \\
2.06465		&	0.60	&		 11.08027	&		11.10073	&		5.00266  \\
2.06465		&	0.70	&		 11.39669	&		11.33588	&		4.80674  \\
2.06465		&	0.80	&		 11.69174	&		11.55846	&		4.65419  \\
2.06465		&	0.90	&		 11.96991	&		11.76663	&		4.53392  \\
2.06465		&	1.00	&		 12.23437	&		11.96089	&		4.43538  \\
2.06465		&	1.10	&		 12.48741	&		12.14250	&		4.35065  \\
2.06465		&	1.20	&		 12.73083	&		12.31285	&		4.27440  \\
2.06465		&	1.30	&		 12.96602	&		12.47327	&		4.20327  \\
2.56465		&	0.10	&		 8.59818	&		10.35536	&		6.81650  \\
2.56465		&	0.20	&		 9.37536	&		10.38058	&		6.71214  \\
2.56465		&	0.30	&		 9.91516	&		10.47251	&		6.59233  \\
2.56465		&	0.40	&		 10.35484	&		10.64227	&		6.45354  \\
2.56465		&	0.50	&		 10.73692	&		10.86349	&		6.29013  \\
\hline
\enddata
\label{tab:COEOS}
\end{deluxetable}

\begin{deluxetable}{ccccc}
	\tablecolumns{5}
	\tablewidth{0pt}
	\tablenum 2
	\tablecaption{Equation of state for CO continued}
	\tablehead{
		\colhead {$\log T$}    & \colhead{$\log \rho$}  & \colhead{$\log P$}  & \colhead{$\log u$} & \colhead{$\log s$}  \\
		\colhead{$[K]$}  & \colhead{$[g/cc]$} & \colhead{$[Pa]$}  & \colhead{$[erg/g]$} & \colhead{$[erg/g-K]$} }
	\startdata
2.56465		&	0.60	&		 11.08076	&		11.10182	&		6.11803  \\
2.56465		&	0.70	&		 11.39686	&		11.33629	&		5.91384  \\
2.56465		&	0.80	&		 11.69180	&		11.55861	&		5.70071  \\
2.56465		&	0.90	&		 11.96994	&		11.76669	&		5.48884  \\
2.56465		&	1.00	&		 12.23438	&		11.96091	&		5.29094  \\
2.56465		&	1.10	&		 12.48742	&		12.14251	&		5.11592  \\
2.56465		&	1.20	&		 12.73083	&		12.31286	&		4.96608  \\
2.56465		&	1.30	&		 12.96602	&		12.47328	&		4.83886  \\
3.06465		&	0.10	&		 8.96463	&		10.41261	&		7.05238  \\
3.06465		&	0.20	&		 9.49348	&		10.43930	&		7.00870  \\
3.06465		&	0.30	&		 9.95968	&		10.52059	&		6.94781  \\
3.06465		&	0.40	&		 10.37374	&		10.67408	&		6.87899  \\
3.06465		&	0.50	&		 10.74579	&		10.88182	&		6.80456  \\
3.06465		&	0.60	&		 11.08515	&		11.11162	&		6.72332  \\
3.06465		&	0.70	&		 11.39912	&		11.34141	&		6.63379  \\
3.06465		&	0.80	&		 11.69298	&		11.56126	&		6.53419  \\
3.06465		&	0.90	&		 11.97055	&		11.76803	&		6.42218  \\
3.06465		&	1.00	&		 12.23471	&		11.96163	&		6.31293  \\
3.06465		&	1.10	&		 12.48758	&		12.14285	&		6.16560  \\
3.06465		&	1.20	&		 12.73092	&		12.31302	&		6.00969  \\
3.06465		&	1.30	&		 12.96606	&		12.47335	&		5.84475  \\
3.56465		&	0.10	&		 9.36962	&		10.57254	&		7.21587  \\
3.56465		&	0.20	&		 9.69405	&		10.59071	&		7.18659  \\
3.56465		&	0.30	&		 10.06369	&		10.65281	&		7.15510  \\
3.56465		&	0.40	&		 10.43076	&		10.77475	&		7.12114  \\
3.56465		&	0.50	&		 10.77783	&		10.94870	&		7.08214  \\
3.56465		&	0.60	&		 11.10251	&		11.15104	&		7.03402  \\
3.56465		&	0.70	&		 11.40902	&		11.36430	&		6.98398  \\
3.56465		&	0.80	&		 11.69886	&		11.57468	&		6.93149  \\
3.56465		&	0.90	&		 11.97416	&		11.77606	&		6.87593  \\
3.56465		&	1.00	&		 12.23696	&		11.96647	&		6.81651  \\
3.56465		&	1.10	&		 12.48903	&		12.14584	&		6.75230  \\
3.56465		&	1.20	&		 12.73184	&		12.31488	&		6.68219  \\
3.56465		&	1.30	&		 12.96665	&		12.47450	&		6.60487  \\
4.06465		&	-1.00	&		 8.72672	&		11.12752	&		7.61469  \\
4.06465		&	-0.90	&		 8.84338	&		11.11787	&		7.60037  \\
4.06465		&	-0.80	&		 8.96101	&		11.10873	&		7.58527  \\
4.06465		&	-0.70	&		 9.07819	&		11.09997	&		7.56926  \\
4.06465		&	-0.60	&		 9.19305	&		11.09161	&		7.55219  \\
4.06465		&	-0.50	&		 9.30353	&		11.08214	&		7.53390  \\
4.06465		&	-0.40	&		 9.40769	&		11.07207	&		7.51424  \\
4.06465		&	-0.30	&		 9.50433	&		11.06010	&		7.49310  \\
4.06465		&	-0.20	&		 9.59411	&		11.04556	&		7.47045  \\
4.06465		&	-0.10	&		 9.68161	&		11.02824	&		7.44633  \\
4.06465		&	0.00	&		 9.77804	&		11.00927	&		7.42091  \\
4.06465		&	0.10	&		 9.90243	&		10.99228	&		7.39448  \\
4.06465		&	0.20	&		 10.07443	&		10.98353	&		7.36684  \\
\hline
\enddata
\end{deluxetable}

\begin{deluxetable}{ccccc}
	\tablecolumns{5}
	\tablewidth{0pt}
	\tablenum 2
	\tablecaption{Equation of state for CO continued}
	\tablehead{
		\colhead {$\log T$}    & \colhead{$\log \rho$}  & \colhead{$\log P$}  & \colhead{$\log u$} & \colhead{$\log s$}  \\
		\colhead{$[K]$}  & \colhead{$[g/cc]$} & \colhead{$[Pa]$}  & \colhead{$[erg/g]$} & \colhead{$[erg/g-K]$} }
	\startdata
4.06465		&	0.30	&		 10.30242	&		10.99646	&		7.33856  \\
4.06465		&	0.40	&		 10.57225	&		11.04588	&		7.30967  \\
4.06465		&	0.50	&		 10.86205	&		11.14146	&		7.28014  \\
4.06465		&	0.60	&		 11.15517	&		11.27981	&		7.24992  \\
4.06465		&	0.70	&		 11.44272	&		11.44692	&		7.21887  \\
4.06465		&	0.80	&		 11.72104	&		11.62736	&		7.18684  \\
4.06465		&	0.90	&		 11.98825	&		11.80866	&		7.14860  \\
4.06465		&	1.00	&		 12.24611	&		11.98690	&		7.10900  \\
4.06465		&	1.10	&		 12.49512	&		12.15893	&		7.06828  \\
4.06465		&	1.20	&		 12.73600	&		12.32343	&		7.02608  \\
4.06465		&	1.30	&		 12.96954	&		12.48020	&		6.98198  \\
4.56465		&	-1.00	&		 9.39391	&		11.85300	&		7.82194  \\
4.56465		&	-0.90	&		 9.49858	&		11.83884	&		7.80733  \\
4.56465		&	-0.80	&		 9.60479	&		11.82481	&		7.79226  \\
4.56465		&	-0.70	&		 9.71228	&		11.81091	&		7.77665  \\
4.56465		&	-0.60	&		 9.82054	&		11.79708	&		7.76038  \\
4.56465		&	-0.50	&		 9.92889	&		11.78321	&		7.74336  \\
4.56465		&	-0.40	&		 10.03655	&		11.76908	&		7.72545  \\
4.56465		&	-0.30	&		 10.14277	&		11.75438	&		7.70651  \\
4.56465		&	-0.20	&		 10.24707	&		11.73869	&		7.68639  \\
4.56465		&	-0.10	&		 10.34964	&		11.72211	&		7.66492  \\
4.56465		&	0.00	&		 10.45180	&		11.70292	&		7.64198  \\
4.56465		&	0.10	&		 10.55668	&		11.68314	&		7.61745  \\
4.56465		&	0.20	&		 10.66971	&		11.66302	&		7.59129  \\
4.56465		&	0.30	&		 10.79841	&		11.64539	&		7.56357  \\
4.56465		&	0.40	&		 10.95060	&		11.63526	&		7.53444  \\
4.56465		&	0.50	&		 11.13044	&		11.64026	&		7.50411  \\
4.56465		&	0.60	&		 11.33652	&		11.66954	&		7.47311  \\
4.56465		&	0.70	&		 11.56221	&		11.72900	&		7.44136  \\
4.56465		&	0.80	&		 11.79910	&		11.82012	&		7.40932  \\
4.56465		&	0.90	&		 12.04056	&		11.93746	&		7.37731  \\
4.56465		&	1.00	&		 12.28172	&		12.07201	&		7.34534  \\
4.56465		&	1.10	&		 12.51981	&		12.21535	&		7.31336  \\
4.56465		&	1.20	&		 12.75346	&		12.36132	&		7.28123  \\
4.56465		&	1.30	&		 12.98189	&		12.50572	&		7.24694  \\
5.06465		&	-2.20	&		 9.01221	&		12.79781	&		8.22267  \\
5.06465		&	-2.10	&		 9.10356	&		12.78341	&		8.20949  \\
5.06465		&	-2.00	&		 9.19496	&		12.76882	&		8.19617  \\
5.06465		&	-1.90	&		 9.28643	&		12.75405	&		8.18271  \\
5.06465		&	-1.80	&		 9.37804	&		12.73910	&		8.16910  \\
5.06465		&	-1.70	&		 9.46983	&		12.72399	&		8.15535  \\
5.06465		&	-1.60	&		 9.56187	&		12.70871	&		8.14144  \\
5.06465		&	-1.50	&		 9.65423	&		12.69328	&		8.12737  \\
5.06465		&	-1.40	&		 9.74698	&		12.67771	&		8.11312  \\
5.06465		&	-1.30	&		 9.84023	&		12.66200	&		8.09870  \\
5.06465		&	-1.20	&		 9.93407	&		12.64619	&		8.08407  \\
5.06465		&	-1.10	&		 10.02860	&		12.63027	&		8.06923  \\
\hline
\enddata
\end{deluxetable}

\begin{deluxetable}{ccccc}
	\tablecolumns{5}
	\tablewidth{0pt}
	\tablenum 2
	\tablecaption{Equation of state for CO continued}
	\tablehead{
		\colhead {$\log T$}    & \colhead{$\log \rho$}  & \colhead{$\log P$}  & \colhead{$\log u$} & \colhead{$\log s$}  \\
		\colhead{$[K]$}  & \colhead{$[g/cc]$} & \colhead{$[Pa]$}  & \colhead{$[erg/g]$} & \colhead{$[erg/g-K]$} }
	\startdata
5.06465		&	-1.00	&		 10.12395	&		12.61427	&		8.05415  \\
5.06465		&	-0.90	&		 10.22023	&		12.59820	&		8.03880  \\
5.06465		&	-0.80	&		 10.31750	&		12.58210	&		8.02315  \\
5.06465		&	-0.70	&		 10.41581	&		12.56597	&		8.00716  \\
5.06465		&	-0.60	&		 10.51516	&		12.54983	&		7.99078  \\
5.06465		&	-0.50	&		 10.61545	&		12.53369	&		7.97396  \\
5.06465		&	-0.40	&		 10.71652	&		12.51752	&		7.95663  \\
5.06465		&	-0.30	&		 10.81818	&		12.50129	&		7.93872  \\
5.06465		&	-0.20	&		 10.92019	&		12.48494	&		7.92015  \\
5.06465		&	-0.10	&		 11.02242	&		12.46839	&		7.90082  \\
5.06465		&	0.00	&		 11.12490	&		12.45155	&		7.88063  \\
5.06465		&	0.10	&		 11.22793	&		12.43435	&		7.85947  \\
5.06465		&	0.20	&		 11.33226	&		12.41683	&		7.83721  \\
5.06465		&	0.30	&		 11.43918	&		12.39913	&		7.81373  \\
5.06465		&	0.40	&		 11.55056	&		12.38222	&		7.78890  \\
5.06465		&	0.50	&		 11.66886	&		12.36618	&		7.76260  \\
5.06465		&	0.60	&		 11.79684	&		12.35360	&		7.73472  \\
5.06465		&	0.70	&		 11.93708	&		12.34707	&		7.70521  \\
5.06465		&	0.80	&		 12.09132	&		12.35034	&		7.67410  \\
5.06465		&	0.90	&		 12.26000	&		12.36788	&		7.64155  \\
5.06465		&	1.00	&		 12.44186	&		12.40353	&		7.60770  \\
5.06465		&	1.10	&		 12.63451	&		12.45973	&		7.57309  \\
5.06465		&	1.20	&		 12.83475	&		12.53549	&		7.53795  \\
5.06465		&	1.30	&		 13.03937	&		12.62723	&		7.50218  \\
\hline
\enddata
\end{deluxetable}
\section{ACKNOWLEDGEMENTS}

The authors wish to thank Gilles Chabrier and an anonymous referee for many constructive comments.  M.P. is supported by a grant from the Pazy Fund of the Israel Atomic Energy Commission. A.L. is supported by a grant from the Simons Foundation (SCOL \#290360 to D.S.).
The computations for this paper were run on the Odyssey cluster supported by the FAS Division of Science, Research Computing Group at Harvard University. A.L. is grateful to the administrative staff for their technical support. A.V. acknowledges support from ISF grants 770/21 and 773/21.

\bibliographystyle{aasjournal}
\bibliography{amit}

\end{document}